\documentclass[authoryear,preprint,12pt]{elsarticle}

\usepackage{amssymb}

\journal{JQSRT}

\begin{document}

\begin{frontmatter}



\title{Symmetry and the generation and measurement of optical torque}


\author{Timo A. Nieminen}
\ead{timo@physics.uq.edu.au}
\author{Theodor Asavei}
\author{Vincent L. Y. Loke}
\author{Norman R. Heckenberg}
\author{Halina Rubinsztein-Dunlop}

\address{Centre for Biophotonics and Laser Science,
School of Physical Sciences,
The University of Queensland,
Brisbane QLD 4072, Australia}

\begin{abstract}
A key element in the generation of optical torque in optical traps,
which occurs when electromagnetic angular momentum is transferred
from the trapping beam to the trapped particle by scattering,
is the symmetries of the scattering particle and the trapping beam.
We discuss the effect of such symmetries on the generation and
measurement of optical torque in optical tweezers, and some
consequent general principles for the design of optically-driven
micromachines.
\end{abstract}


\end{frontmatter}


\section{Introduction}

Optical trapping, exemplified by the single-beam gradient trap,
or \emph{optical tweezers} \citep{ashkin1986}, has developed
from an interesting novelty into a widely-used and powerful tool
with many applications in biophysics and other areas
\citep{ashkin2000,moffitt2008}. Uses include the measurement
of forces on the order of piconewtons, such as the adhesion
forces on a cellular level \citep{knoner2006e} and forces
in single-molecule studies \citep{williams2002}, as well
as manipulation of microscopic objects, including live
bacteria and eukaryote cells.

An optical tweezers trap is conceptually simple: a high
numerical aperture lens, typically a microscope objective
lens, is used to focus a laser beam to a diffraction-limited
spot. The sample, consisting of the microscopic objects one
intends to trap or manipulate (often polystyrene or silica microspheres)
suspended in a fluid, which
would usually be water or a biological buffer solution,
is placed on a stage, and the objects, as long as their
refractive index is higher than that of the suspending
fluid, and they are not too reflective or absorbing
\citep{stilgoe}, can be trapped in the bright focal spot.

The optical forces responsible for trapping are fundamentally
the result of the transfer of momentum from the beam through
scattering of the trapping beam
\citep{ashkin1986,ashkin1992,nieminen}; this possible due
to the transport of momentum by electromagnetic fields,
with the optical force being equal to the rate of transfer
of momentum from the beam to the particle.

Since angular momentum can also be carried by electromagnetic
fields, including light, the possibility of exerting optical
torque on particles, trapped or otherwise, is evident.
This introduces the possibility of six-degrees-of-freedom,
with the orientation of a trapped particle being controlled,
in addition to its position---this would be a major advance
in the manipulation possible with optical tweezers.
Rotation can also be used as a tool to actively
probe the physical properties of the surrounding medium
\citep{nieminen2001jmo,bishop2004,laporta2004}.
In addition, it could allow the development of practical
optically-driven micromachines
\citep{friese2001,galajda2001,nieminen2006a}.
A significant amount of progress has been achieved in 
all of these areas
\citep[e.g.,][]{bayoudh2003,parkin2007b,kelemen2006},
using a diverse range of methods to generate the optical
torque.

A key element determining whether or not the transfer
of electromagnetic angular momentum to a particle by scattering
is possible is the symmetries of the scattering particle and
the trapping beam
\citep{konz2003,nieminen2004a,benford2004,nieminen2004c}.
Therefore, consideration of the effects of symmetry can provide
insight into rotational optical micromanipulation and
design of optically-driven micromachines and a unified overview
of the generation of optical torque.

\section{Optical angular momentum}

Optical forces have a long and interesting history \citep{worrall1982},
starting with Kepler's conjecture that radiation pressure was responsible
to the direction of comet tails \citep{kepler1619}. However,
due to their smallness, practical
applications in the laboratory did not appear feasible until
Ashkin noted that only very small forces are needed to manipulate
very small objects \citep{ashkin2000}, which, notably, led to the
invention of optical tweezers \citep{ashkin1986}. The
fundamental principle by which optical tweezers operate is the
exchange of momentum between the trap beam and a particle
within the trap \citep{ashkin1986,nieminen2007d}.

It was suggested as early as 1909 by \citet{poynting1909} that light can
carry angular momentum as linear momentum. While
Poynting was skeptical as to whether such torques could be
measured, this was achieved in
1936 by \citet{holbourn1936} and \citet{beth1936}.

Most of the earliest optical tweezers-based
demonstrations of optical torque made use of absorption, combined
with laser beams carrying angular momentum
\citep{he1995,friese1996pra,simpson1997}, which has very limited
practical use due to the resulting heating, which, for example,
could lead to rapid death of live specimens, and in any case,
makes three-dimensional trapping more difficult or even impossible.

More promising was the use of birefringent materials
\citep{friese1998nature,bishop2004,laporta2004},
or specially-shaped objects
\citep{higurashi1994,higurashi1997,bonin2002,cheng2002,bishop2003,neale2005}
which can themselves be made of birefringent materials,
or be shape-birefringent \citep{born1997}.

Notably, these types of methods, using birefringence (due to
either the material or shape) which affects the polarization
of the trapping beam and therefore depends on the
\emph{spin} angular momentum of light, and the use of shapes
which 
depend on, and affect, the spatial structure of the trapping
beam, and therefore make use of the \emph{orbital} angular
momentum of light, are often viewed as distinct
physical mechanisms.

It is true that there are important differences between
spin and orbital angular momenta, but few of them will
be of importance here. One difference that will turn out
to be important is that spin angular momentum can be readily
determined by measurement of the Stokes parameters
\citep{crichton2000}. However, it is still worth reviewing
the basic elements of electromagnetic angular momentum, especially
since there is ongoing debate within the literature concerning
the angular momentum of circularly polarized waves
\citep{humblet1943,stewart2005a,pfeifer2006c}.

The controversy arises since there are two paths we can take
to determining the angular momentum density or flux.
Firstly, we can assume that the angular
momentum density is the moment of the momentum density,
$\mathbf{r}\times(\mathbf{E}\times\mathbf{H})/c$,
where $\mathbf{r}$ is the position vector relative to the
origin of our chosen coordinate system, and
$\mathbf{E}$ and $\mathbf{H}$ are the electric and
magnetic fields respectively, which leads to the curious
result that a circularly polarized plane wave carries
zero angular momentum in the direction
of propagation, in contradiction to the usual quantum mechanical
result of $\pm\hbar$ angular momentum per photon. This choice
of angular momentum density implicitly states that there
is no spin angular momentum---the spin angular momentum is
the part of the total angular momentum which is independ of the
choice of origin about which moments are taken; taking
$\mathbf{r} = 0$ in the above expression for the
angular momentum density yields a spin of zero.

Secondly, we can begin with the Lagrangian
for the electromagnetic radiation field, and, via
Noether's (\citeyear{noether1918}) theorem, obtain the canonical
stress tensor and an angular momentum tensor. The angular momentum
tensor can be divided into
spin and orbital components \citep{jauch1976,soper1976},
based on the independence
of the spin from the choice of origin. For a circularly polarized
wave, the spin would be $I/\omega$, where $I$ is the irradiance and
$\omega$ the angular frequency, in disagreement with the first
result, and in agreement with the quantum mechanical result.
This separation of the angular momentum density
into spin and orbital parts is, in general, not
gauge-invariant, and it is common to transform the canonical
stress tensor into a summetric stress tensor, with a
gauge-invariant result for the angular momentum density, yielding
the integral of
$\mathbf{r}\times(\mathbf{E}\times\mathbf{H})/c$.
\citet{jauch1976} carefully point out that this transformation
requires the dropping of surface terms at infinity. The reverse of this
procedure, obtaining the spin and orbital terms starting from
$\mathbf{r}\times(\mathbf{E}\times\mathbf{H})/c$,
involving the same surface terms, had already been shown
by \citet{humblet1943}. For and physically realisable beam,
these surface terms vanish, and despite the controversy,
both expressions for the angular momentum density yield the
same total angular momentum and torque exerted on
an object \citep{zambrini2005a,nieminen2007b}.

Since we are interested in torque within optical tweezers,
where we have a monochromatic wave, the gauge-invariance
of the separation of the angular momentum
into spin and orbital parts is also a non-issue, since in
this case, the separation is gauge-invariant
\citep{crichton2000,barnett2002}.

Consequently, the spin angular momentum flux of a paraxial
electromagnetic beam, where the wavefronts are almost plane
(by definition), varies from $-\hbar$ per photon in the
direction of propagation for right circular polarization,
through zero for linear polarization, to $+\hbar$ per photon
for left circular polarization. In classical terms,
these extremes are $\pm P/\omega$, where $P$ is the beam power.
Linear or elliptical
polarization can be represented by the superposition of two
circularly polarized components of opposite handedness; the
total spin angular momentum is equal to the sum of the momentum
of the two components, which, if they are of equal power, will
be zero.

Since, in the paraxial limit, the spatial structure of the
beam is independent of the polarization (since it is described
by a solution of the \emph{scalar} paraxial wave equation),
the spin and orbital angular momenta are independent of
each other \citep{allen1992,allen1999b}. 
Thus, an excellent formalism for the description of the angular
momentum of paraxial beams is provided by the combination
of Laguerre--Gauss beam modes and circularly polarized
components \citep{heckenberg1999,allen1999b,nienhuis}.
Each Laguerre--Gauss mode is described by a radial mode
index $p$ and an azimuthal mode index $\ell$, and is
denoted LG$_{p\ell}$. Each mode has an azimuthal phase
dependence of $\exp(\mathrm{i}\ell\phi)$, where $\phi$ is
the azimuthal angle. This azimuthal dependence is typical
of that in solutions to PDEs in rotational coordinate systems
obtained by separation of variables. Each mode also has a well-defined
orbital angular momentum flux of $\ell\hbar$ per photon about the
beam axis, or $\ell P/\omega$, which is also typical of such
solutions. In this case, the modes are orthogonal w.r.t. orbital
angular momentum, and the total orbital angular momentum about the
beam axis, $L_z$, can
be found by adding the individual angular momenta of each mode:
\begin{equation}
L_z = \frac{1}{\omega} \sum_{p=0}^\infty \sum_{\ell = -\infty}^\infty
\ell |a_{p\ell}|^2,
\end{equation}
where $a_{p\ell}$ are the mode amplitudes, and units are
chosen such that the total power is equal to 
\begin{equation}
P = \sum_{p=0}^\infty \sum_{\ell = -\infty}^\infty |a_{p\ell}|^2.
\end{equation}
The total angular momentum flux about the beam axis, $J_z$, is
then the sum of the orbital ($L_z$) and spin ($S_z$) components,
\begin{equation}
J_z = L_z + S_z,
\end{equation}
where
\begin{equation}
S_z = \sigma_z P/\omega,
\end{equation}
where $\sigma_z$ is the degree of circular polarization, which
varies from $-1$ for right circular polarization to $+1$ for
left circular.

However, for a tightly focussed beam such as employed in optical
tweezers, the paraxial approximation cannot be used---it is necessary
to use modes which are solutions of the vector Helmholtz equation.
It is convenient to begin with a set of solutions to the scalar
Helmholtz equation, $\psi_n$, from which a set of solutions to
the vector Helmholtz equation can be obtained:
\begin{eqnarray}
\mathbf{L}_n & = & \nabla \psi_n, \\
\mathbf{M}_n & = & - \hat{\mathbf{a}} \times \mathbf{L}_n,\\
\mathbf{N}_n & = & \frac{1}{k} \nabla\times\mathbf{M}_n,
\end{eqnarray}
where $\hat{\mathbf{a}}$ is a unit vector or a constant
vector \citep{brock2001}.
The $\mathbf{L}_n$ wavefunctions are curl-free, while the $\mathbf{M}_n$
and $\mathbf{N}_n$ wavefunctions are divergence-free. Therefore, for
electromagnetic fields, we only require $\mathbf{M}_n$ and $\mathbf{N}_n$.
We can also note that $\mathbf{M}_n = (1/k) \nabla\times\mathbf{N}_n$.

In spherical coordinates, we can readily write a
general solution to the scalar Helmholtz equation:
\begin{equation}
\psi = \sum_{n=0}^\infty \sum_{m=-n}^{n}
a_{nm} j_n(kr) Y_{nm}
\end{equation}
where $a_{nm}$ are the mode amplitudes, $j_n$ are
spherical Bessel functions, and
$Y_{nm}$ are normalised spherical harmonics. The resulting
vector solutions are the regular vector multipole fields,
or vector spherical wavefunctions (VSWFs):
\begin{eqnarray}
\mathbf{M}_{nm}(k\mathbf{r}) & = & N_n j_n(kr)
\mathbf{C}_{nm}(\theta,\phi) \\
\mathbf{N}_{nm}(k\mathbf{r}) & = & \frac{j_n(kr)}{krN_n}
\mathbf{P}_{nm}(\theta,\phi) + N_n
\left( j_{n-1}(kr) -
\frac{n j_n(kr)}{kr} \right) \mathbf{B}_{nm}(\theta,\phi)
\nonumber
\end{eqnarray}
where 
$N_n = [n(n+1)]^{-1/2}$ are normalization constants, and
$\mathbf{B}_{nm}(\theta,\phi) = \mathbf{r} \nabla Y_n^m(\theta,\phi)$,
$\mathbf{C}_{nm}(\theta,\phi) = \nabla \times \left( \mathbf{r}
Y_n^m(\theta,\phi) \right)$, and
$\mathbf{P}_{nm}(\theta,\phi) = \hat{\mathbf{r}} Y_n^m(\theta,\phi)$
are the vector spherical
harmonics
\citep{mishchenko2000book,waterman1971,mishchenko1991,jackson1999c}.

Since the spherical harmonics $Y_{nm}$ have an azimuthal dependence
of $\exp(\mathrm{i}m\phi)$, we once again obtain angular
momentum about the beam axis in terms of the azimuthal mode
index ($m$); in this case we have the \emph{total} angular momentum
about the beam axis equal to $m\hbar$ per photon, or $mP/\omega$,
rather than just the orbital angular momentum. The radial mode index
$n$ also relates to the angular momentum---the VSWFs
are eigenfunctions of the angular momentum operator
$J^2$, with eigenvalues $[n(n+1)]^{1/2}$, as well as the angular
momentum operator $J_z$, with
eigenvalues $m$. As a consequence of this, it is clear that we
must have $n\ge|m|$. The radial behaviour of each VSWF mode
depends on spherical Bessel functions $j_n(kr)$, which are peaked
in the vicinity of $kr\approx n$. Therefore, the minimum width of
a focussed beam depends on its angular momentum flux \citep{nieminen3};
this has been noted both experimentally \citep{curtis2003b} and
on elementary theoretical grounds \cite{courtial2000}.
An interesting result of this, since the minimum width of a beam
also depends on its wavelength, which also affects the angular
momentum flux due to the dependence of angular momentum on the
frequency (or in quantum terms, the angular momentum per photon
is constant, and the photon flux increases as the frequency is
decreased), this minimum width is independent of the wavelength
and whether the angular momentum is spin, orbital, or a combination
of the two.

Since spherical harmonics are a common feature of solutions to PDEs
obtained by separation of variables in spherical coordinate systems,
they, and their relationship to angular momentum, are discussed in
many sources, most commonly in the context of quantum mechanics
\citep[e.g.,][]{rose,varshalovich}.

\section{Symmetry, scattering, and angular momentum}

Since we are already considering the trapping beam as a superposition
of VSWFs, it is natural to use the T-matrix formalism
\citep{nieminen2003b,kahnert2003b,mishchenko2000book,waterman1971,%
mishchenko1991,jackson1999c}. For this, we need to be able to describe
the scattered wave, for which we use the outgoing VSWFs instead of
the regular VSWFs; to obtain these, we simply replace the spherical
Bessel functions $j_n(kr)$ in the regular VSWFs with spherical Hankel
functions of the first kind $h_n^{(1)}(kr)$ (assuming a time-dependence
of $\exp(-\mathrm{i}\omega t)$---if we have chosen a time dependence
of  $\exp(\mathrm{i}\omega t)$, we use spherical Hankel
functions of the second kind $h_n^{(2)}(kr)$).

Since ``T-matrix method'' is often used synonymously with ''extended
boundary condition method'' (EBCM) which, properly speaking, is a method
of calculating the T-matrix \citep{nieminen2003b,kahnert2003b}, it is worth
briefly describing the \textit{T}-matrix description of scattering.
Essentially, we expand both 
the incident and scattered waves in terms of a sufficiently
complete basis set of functions ($\psi_n^{(\mathrm{inc})}$ for the incident
wave, and $\psi_n^{(\mathrm{scat})}$ for the scatterered wave, where
$n$ is a mode index labelling the discrete modes), each of which is a
solution of the Helmholtz equation:
\begin{equation}
U_\mathrm{inc} = \sum_n^\infty a_n \psi_n^{(\mathrm{inc})},
\label{exp1}
\end{equation}
\begin{equation}
U_\mathrm{scat} = \sum_k^\infty p_k \psi_k^{(\mathrm{scat})},
\label{exp2}
\end{equation}
where $a_n$ are the expansion coefficients for the incident wave,
and $p_k$ are the expansion coefficients for the scattered wave.
If the response of the scatterer is linear, the relationship between the
incident and scattered waves can be written as a linear system
\begin{equation}
p_k = \sum_n^\infty T_{kn} a_n
\end{equation}
or, in more concise notation, as the matrix equation
\begin{equation}
\mathbf{P} = \mathbf{T} \mathbf{A}
\end{equation}
where $T_{kn}$ are the elements of the \textit{T}-matrix.

Note that apart from the assumptions that  the scattering is elastic
and monochromatic, there are no other restrictions other than that
 a suitable basis set exists.
This formalism can be used for a wide range geometries, including
compact particles (ideally treated in spherical coordinates), generalised
cylindrical particles and surfaces, and different types of waves, including scalar
and vector waves.

Since we are interested in the transfer of angular momentum to the scatterer,
it is convenient to make use a basis of incoming and outgoing VSWFs, rather
than regular and outgoing. The incoming VSWFs are simply
obtained by replacing the spherical
Bessel functions $j_n(kr)$ in the regular VSWFs with spherical Hankel
functions of the second kind $h_n^{(2)}(kr)$.

Thus, while we began with the incident field written as
\begin{equation}
\mathbf{E}_\mathrm{inc}(\mathrm{r}) = \sum_{n=1}^\infty \sum_{m = -n}^n
a^{(3)}_{nm} \mathbf{M}_{nm}(k\mathrm{r}) +
b^{(3)}_{nm} \mathbf{N}_{nm}(k\mathrm{r}),
\label{regular_expansion}
\end{equation}
we can replace this with the purely incoming part of the incident field,
\begin{equation}
\mathbf{E}_\mathrm{inc}(\mathrm{r}) = \sum_{n=1}^\infty \sum_{m = -n}^n
a^{(2)}_{nm} \mathbf{M}_{nm}^{(2)}(k\mathrm{r}) +
b^{(2)}_{nm} \mathbf{N}_{nm}^{(2)}(k\mathrm{r}).
\label{incoming_expansion}
\end{equation}
In both cases, the scattered field is written
\begin{equation}
\mathbf{E}_\mathrm{scat}(\mathrm{r}) = \sum_{n=1}^\infty \sum_{m = -n}^n
p^{(1)}_{nm} \mathbf{M}_{nm}^{(1)}(k\mathrm{r}) +
q^{(1)}_{nm} \mathbf{N}_{nm}^{(1)}(k\mathrm{r}),
\label{outgoing_expansion}
\end{equation}
but it is important to note that the ``scattered'' field is not the
same in the two cases---in the first case, it is the scattered field
in the usual usage of the term, while in the second, it is the total outgoing
portion of the field, which includes the outgoing part of the incident
field. Since $j_n = ( h_n^{(1)} + h_n^{(2)})/2$,
$a^{(3)}_{nm} = 2a^{(2)}_{nm}$ and $b^{(3)}_{nm} = 2b^{(2)}_{nm}$, and
the purely outgoing mode coefficients for the second case
($a_{nm}^{(1),\mathrm{out}}, b^{(1),\mathrm{out}}$) are related to the
scattered field amplitudes in the first case
($a_{nm}^{(1),\mathrm{scat}}, b^{(1),\mathrm{scat}}$) by
$a_{nm}^{(1),\mathrm{out}} = a^{(3)}_{nm}/2 + a_{nm}^{(1),\mathrm{scat}}$
and
$b_{nm}^{(1),\mathrm{out}} = b^{(3)}_{nm}/2 + b_{nm}^{(1),\mathrm{scat}}$.
The T-matrix itself will also differ in the two cases---using
the regular expansion, the
\textit{T}-matrix in the absence of a scatterer is a zero matrix, while using
the incoming field expansion, the no-scatterer \textit{T}-matrix is the identity
matrix. The relationship between them is $\mathbf{T}^{\mathrm{(in/out)}} =
2\mathbf{T}^{\mathrm{(inc/scat)}} + \mathbf{I}$; the T-matrix for the second
case is typically written as $\mathbf{S}$ and called the S-matrix.

The angular momentum flux into or out of the system can be found by
integrating the moment of the momentum flux density
\begin{equation}
\mathbf{j} = \frac{1}{2c}
\mathrm{Re}(\mathbf{r}\times(\mathbf{E}\times\mathbf{H}^\ast))
\end{equation}
over a sphere surrounding the scatterer. Taking care to find the
moment first, and then taking the far-field limit, this integral
can be performed analytically if the fields are written in terms of VSWFs,
for normalised angular momentum about the $z$-axis of
\begin{equation}
\tau_z = \sum_{n=1}^\infty \sum_{m = -n}^n m
( |a_{nm}|^2 + |b_{nm}|^2
- |p_{nm}|^2 - |q_{nm}|^2 ) / P
\label{torque}
\end{equation}
where
\begin{equation}
P = \sum_{n=1}^\infty \sum_{m = -n}^n |a_{nm}|^2 + |b_{nm}|^2
\end{equation}
is the incident power. This is the optical torque exerted
on the scatterer, in units of $\hbar$ per photon.
The spin and orbital contributions to the torque
can be calculated if desired \citep{crichton2000,bishop2003}; the
spin torque is
\begin{eqnarray}
\sigma_z & = & \frac{1}{P} \sum_{n=1}^\infty \sum_{m = -n}^n
\frac{m}{n(n+1)} 
( |a_{nm}|^2 + |b_{nm}|^2 - |p_{nm}|^2 - |q_{nm}|^2)
\nonumber \\ & & - \frac{2}{n+1}
\left[ \frac{n(n+2)(n-m+1)(n+m+1)}{(2n+1)(2n+3)} \right]^{\frac{1}{2}}
\nonumber\\ & &
\times \mathrm{Im}( a_{nm} b_{n+1,m}^\star + b_{nm} a_{n+1,m}^\star
- p_{nm} q_{n+1,m}^\star - q_{nm} p_{n+1,m}^\star ).
\end{eqnarray}
The remainder of the torque is the orbital contribution.

In optical tweezers, the incident beam that is focussed by the
objective to produce the trapping beam typically has uniform orbital
angular momentum. In conventional optical trapping, this is a Gaussian
beam, with zero orbital angular momentum, while if one is intending
to rotate trapped particles, this could be a Laguerre--Gauss beam
carrying non-zero orbital angular momentum. In either case,
the beam can also carry spin angular momentum due to circular
polarization. The angular momentum flux of this incident beam is
determined by the Laguerre--Gauss azimuthal mode $\ell$, and the
degree of circular polarization $\sigma_z$. The polarization state
will, in general, be the result of superposition of left- and
right-circular modes with $\sigma_z = \pm 1$.
After focussing, however, the angular momentum is described by
the VSWF azimuthal mode $m$. The total angular momentum is unchanged by
being focussed \citep{nieminen2}, so
\begin{itemize}
\item only $m = \ell \pm 1$ VSWF modes can be nonzero,
\item for circular polarisation, either the $m = \ell - 1$ modes or
the $m = \ell + 1$ modes will be nonzero, but not both,
\item and both will be nonzero for plane or elliptical polarisation.
\end{itemize}
Thus, we will typically have two sets of modes, separated by an
angular momentum difference of $\Delta m = 2$.

While finding the VSWF expansion from the original Laguerre--Gauss
expansion presents both theoretical and practical difficulties, primarily
as a result of the VSWF being solutions to the vector Helmholtz equation,
while the Laguerre--Gauss modes are solutions to the scalar paraxial wave equation,
there are a number of ways in which it can be done
\citep{nieminen2003a,nieminen2007b}.

What then of the effect of the symmetry of the scattering particle?
With respect to the azimuthal coordinate $\phi$, \emph{every} object
is periodic, even if completely devoid of symmetry (in which case, the
period is $2\pi$, since the object must be rotated by a complete
rotation before matching its original appearance). Thus, the azimuthal
dependence of the fields and the boundary conditions (i.e., the effect
of the scatterer) present an exact mathematical analogy with the case
of a planar grating illuminated by a plane wave. For the grating,
if the component of the wavevector of the incident wave along the
grating is $k_{x}$, and the component of the reciprocal lattice
vector of the grating along the grating is $q_x$, the scattered waves
have a discrete angular spectrum with wavevectors given by
$k_{xj} = k_x + jq_x$, where $j$ is the order of scattering. This is
a fundamental result of Floquet's theorem. In the case of the azimuthal
behaviour of the fields, the azimuthal mode index $m$ plays the part of
the $x$ component of the wavevector, and the order of discrete rotational
symmetry $p$ (a scatterer has $p$-th order discrete rotational symmetry
if it is indistinguishable $p$ times during a complete rotation) is
equivalent to the lattice vector. Thus, a discrete azimuthal spectrum,
with azimuthal mode indices
\begin{equation}
m_j = m_0 + jp
\label{coupling}
\end{equation}
is obtained, where $m_0$ is the azimuthal mode index of the incident wave.
Since $m_j$ determines the angular momentum of each mode,
we see that the optical torque is strongly influenced by the
symmetry of the scatterer.

For
scatterers that are mirror-symmetric, upward and downward coupling must be
equal, in the sense that, for example, a mirror-symmetric scatterer of
2nd order rotational symmetry (such as a long rod), T-matrix elements
coupling from $m = 1$ modes to $m = -1$ modes will have the same magnitudes
as the elements coupling from $m = -1$ to $m = 1$ modes. For chiral
scatterers, these T-matrix elements will, in general, be different.

Two cases can be simply dealt with. Firstly, if the particle is devoid
of symmetry (i.e., $p=1$), the incident wave is coupled to all possible
scattered modes, and little can be said about the details of
possible effects. Secondly, if the particle is axisymmetric
($p\rightarrow\infty$), we only have $m_j = m_0$, and the angular momentum
per photon is unchanged. Thus, an axisymmetric particle can only experience
an optical or electromagnetic torque if the photon flux changes due to
absorption (or, in principle, gain). This lack of coupling between
different angular momenta is widely exploited in the calculation
of the T-matrix
\citep{nieminen2003b,mishchenko2000book,waterman1971,mishchenko1991}, in
additions to the benefit of axisymmetry turning 2D surface integrals
into 1D line integrals, or 3D volume problems into 2D surface problems
\citep{loke2007a}. Even in the case of discrete rotational symmetry
rather than axisymmetry, the restricted coupling between modes
(\ref{coupling}) can still be used to reduce computational time
by orders of magnitude \citep{kahnert2005,nieminen2006b,nieminen2007b}.

We can also note some rotation properties of \textit{T}-matrices. Since the
individual VSWF modes have an azimuthal dependence of $\exp(\mathrm{i}m\phi)$,
a rotation of the coordinate system of $\Phi$ about the $z$-axis
must induce a phase change of $m\Phi$ in the VSWF modal
amplitudes. The combination of rotation of the incident and scattered fields
is equivalent to a phase change of $(m_1 - m_2)\Phi$ for the \textit{T}-matrix
element coupling modes of azimuthal index $m_1$ and $m_2$. If the scatterer
is rotating, then $\Phi = \Omega t$, and the scattered light experiences a
rotational Doppler shift of $(m_1 - m_2)\Omega$. This rotational Doppler
shift accounts for the work done by the field on the spinning
particle \citep{atkinson1935}.

Since the angular dependence of the torque
results in alignment to the plane of polarization rather than rotation,
an object can be rotated at a controlled rate by rotating the plane
of polarization. However, it is not possible to control the torque
by changing the degree of circular polarization of a single beam,
since this alignment will prevent spinning about the beam axis
\citep{friese1998nature}. It is possible to overcome this
problem by using two beams, not mutually coherent, with circular
polarization of opposite handedness \citep{maren}, but only at
the cost of a much more complicated trapping apparatus.

While the use of elongated objects or birefringent materials is simple
and can
can allow large torques to be generated \citep{friese1998nature},
there are some drawbacks. Firstly, strongly birefringent materials
such as calcite or vaterite (both calcium carbonate minerals) can be
chemically unsuitable, for example perhaps dissolving rapidly in the
biological buffer solution being used, or not adhering to biomolecules
of interest. The shape birefringence of elongated objects is typically
weak, and the torques obtained are typically a few percent of $\hbar$
per photon \citep{bishop2003}. In addition, elongated object tend to align
along the beam axis, at which point they are axisymmetric or close to
axisymmetric about the beam, and cannot be rotated further. This last
problem can be overcome by using flattened object rather than elongated
ones \citep{cheng2002,bayoudh2003}, but the problem of low torque remains.

\section{$p \le 2$ versus $p>2$: shape birefringence}

If we consider a simple particle such as a mirror-symmetric elongated
particle, we immediately note that such a particle possesses 2nd order
rotational symmetry, and hence couples modes with $m_1 = m_2 \pm 2n$.
We can expect $m_1 = m_2 \pm 2$ to dominate, especially for small particles.

Firstly, we consider the case of such a particle trapped by a linearly polarized
Gaussian beam ($m = \pm 1$) mirror symmetric plane polarized beam.
If the beam is plane polarised along the $x$-axis, the $m = 1$ and $m=-1$
amplitudes are equal. For simplicity, and without loss of generality, we
assume that this is the case.
If the long axis of the particle, and the symmetry axis of the beam coincide,
the coupling from $m = +1$ to $m= -1$ must be identical, and the scattered
modes are of opposing angular momentum are equal. Thus, no torque can be
exerted on the trapped particle. The same considerations apply if the
particle is rotated by 90$^\circ$---in this case, the effect of the rotation
is a multiplication by $-1$ of all relevant \textit{T}-matrix elements, and
the scattered modes of opposing angular momentum are still equal.
If we consider a single VSWF mode of amplitude $a_{n,1}$
incident on the particle, in the absence
of absorption, energy must be conserved, and hence
$p_{n,1} = a_{n,1} (1- \mathrm{i}x)$ and $p_{n,-1} = \mathrm{i}x$ where $x$ is
a real number.

A rotation of 45$^\circ$, however, results in multiplication by $\mathrm{i}$
and $-\mathrm{i}$ of the upward and downward coupling elements, with no change
to the coupling from $m = 1$ to $m= 1$, or $m = -1$ to $m = -1$.
As a result, if the initial mode amplitudes are real, the real part of, say, 
the $m = 1$ mode is increased, and the real part of the $m=-1$ mode is
decreased, while their imaginary parts remain the same. Thus, the scattered
beam now has non-zero angular momentum, and an optical torque is exerted
on the particle.

A left circularly polarized incident beam will simply have the $m = 1$
amplitudes reduced, and the $m= -1$ amplitudes will become non-zero, again
giving rise to an optical torque. In this case, the torque is independent
of the orientation of the particle about the $z$-axis, since the optical
angular momentum is independent of the phase of the mode amplitudes (which
is all that will be changed by rotation of the particle). The same conclusion
can also be simply deduced from the rotational symmetry of the system.

In the case of weak scattering, as will be typical of most small
elongated objects, the coupling will be weak (i.e., $x$ above will be small).
For the circularly polarized beam, this results in loss of power in
$m=1$ modes proportional to $x^2$, and an equal gain in the $m=-1$ modes.
Therefore, the optical torque will be quite small. For a plane polarized
beam at 45$^\circ$ to the particle, the changes in power in each mode
are proportional to $x$, and will generally be much larger. This conclusion
is supported both by rigorous electromagnetic calculations and by experimental
measurements of the torque in the two case \citep{bishop2003}.

Similar considerations apply to the rotation of birefringent particles---these
possess the same symmetry as elongated particles. The major difference is that
strongly birefringent particles will results in stronger coupling (i.e., larger
$x$ above), and the difference noted above between the torque exerted by
circularly and linearly polarized beams need not apply. Energy conservation
considerations give the usual maximum angular momentum transfer of $2\hbar$
per photon for a circularly polarized beam, and $\hbar$ per photon for
a linearly polarized beam. We can note that coupling between the
$m = 1$ and $m = -1$ modes
of equal $n$ is not sufficient to reverse the polarization of a beam since
$1- \mathrm{i}x$ cannot be zero for real $x$---a
waveplate must couple orders of differing $n$ in order to be effective.

If, instead of trapping the elongated particle in a Gaussian beam, we now
use a beam with an elliptical focal spot, the trapping beam has
VSWF modes of $m = \pm 1, \pm 3, \pm 5$, etc, and also modes of higher $n$,
due to the larger extent of the focal spot. Again, the overall behaviour will
be similar to that of an elongated particle trapped in a linearly polarized beam.
For weak coupling, we can expect higher torques, since the outward coupling
to orders of greater $|m|$ now contribute to the torque rather than resulting
in equal angular momenta in the postive and negative $m$ orders. The exact
details, however, are likely to be strongly influenced by the exact particle
geometry. Modes with $m = \pm 1, \pm 3, \pm 5$, etc will generally exist,
due to the overall shape of the beam, regardless of the incident polarization,
and the interference terms that give rise to higher torques in the case
of weak coupling will still exist---a large increase in the torque should
result for circularly polarized beams.

For higher-order rotational symmetry ($p>2$), such angle-dependent
torques are no longer possible.
For example, if an object has 4th-order symmetry, clearly we can write
the components of its polarizability tensor in the plane perpendicular
to the beam axis as $\alpha \mathbf{I}$, where $\alpha$ is a constant
and $\mathbf{I}$ is the $2\times 2$ identity tensor. Due to the
rotational invariance of $\mathbf{I}$, there will be no angle-dependent
torque. The same argument also applies to any higher-order rotational
symmetry with $p>2$. If the scatterer is also mirror-symmetric, there
can be no rotation of the plane of polarization of incident
light, and such a scatterer cannot possess shape birefringence. If
the scatterer is chiral, rotation of the plane of polarization of incident
light, and the scatterer can have circular birefringence (a.k.a. optical
activity). However, circular birefringence will not give rise to
optical torque.

This angle-independence of torque can also be seen in terms of VSWFs.
The two incident circular polarization components are
$\Delta m = 2$ apart in angular momentum, and
a scatterer with $p>2$ cannot couple incident modes with azimuthal
indices $m$ and $m+2$ to the same mode. The scattered modes resulting
from each of the incident polarization components are completely
independent, and the angle-dependence which results from interefence
between modes scattered from each of the two polarization components
cannot occur, since they cannot interfere.

\section{Optically-driven rotors}

One strand of the web of optical rotation has been the practical
implementation of optical rotation of microstructures, especially
specially-fabricated microparticles intended for use as
optically-driven micromotors, micropumps, or other micromachine
components. Work in this field has been recently reviewed
by \citet{nieminen2006a}.

Such optically-driven micromachines typically possess discrete
rotational symmetry, with $p>2$. Therefore, we will consider
some qualitative features of the coupling between angular
momentum orders. Firstly, we can expect scattering to the lowest orders
of scattering (i.e., $n = 0, \pm 1$) to be usually strongest. Thus, most
light is likely to be scattered to scattered wave azimuthal orders
$m = m_0, m_0 \pm p$.

Secondly, only $m$ such that $|m| \le n_\mathrm{max}$ are available, with
$n_\mathrm{max}$ being determined by the radius of a sphere required to
enclose the micromachine element. Therefore, if the order of symmetry $p$
is large, and the micromachine small, only a small number of azimuthal
modes will be available. The ultimate case of this is the homogeneous
isotropic sphere, with $p = \infty$, when only the incident $m = m_0$ is
available for the scattered wave. In principle, this can be exploited
to maximise torque. As a numerical example, consider a
structure with $p = 8$, size such that $n_\mathrm{max} = 6$,
illuminated with a beam such that $m_0 = 4$. In this instance,
the only scattered wave azimuthal orders that are available are
$m = -4, 0, 4$, and we might expect such a structure to generate more
torque, \emph{cetera paribus},
than a structure that can also scatter
to $m = \pm 8$ (since the scattering to $m = +8$ will probably be
stronger than the scattering to $m=-8$). However, such a structure
would have a maximum radius of less than a wavelength, which,
apart from causing fabrication to be difficult, with sub-wavelength
resolution being required even at the perimeter, would mean that
the device would largely sit within the dark centre of a typical
tightly focussed optical vortex. Alternative methods of illumination,
such as a Gaussian or similar beam perpendicular to the device symmetry
axis illuminating one side \citep{kelemen2006}, can avoid this, but will
not necessarily result in greater efficiency.
Thus, designing the particle itself to more efficiently scatter preferentially
into either higher $m$ (i.e., more positive or less negative) or lower $m$
(i.e., less positive or more negative) orders is desirable.

Consequently, it is useful to note that, thirdly,
the symmetry of scattering to positive and negative orders of scattering
(i.e., to $m < m_0$ and $m > m_0$) is dependent of the chirality of the
particle shape. If the particle is achiral---mirror symmetric about a plane
containing the axis of rotational symmetry---the coupling from $m_0 = 0$
to $\pm m$ will be identical, since these modes are mirror-images of each
other. For a chiral particle, however, the coupling
to these mirror-image modes will not be identical; even in this case of
illumination with zero angular momentum modes (i.e., $m_0 = 0$), optical
torque will be generated. If the particle is chiral such that scattering
from $m_0 = 0$ to $+m$ is favoured, then, even for incident modes
with $m_0 \ne 0$, scattering to higher (i.e., more positive) $m$ will
generally be favoured. In this way, a particle can be optimised for
the generation of torque with a particular handedness, leading to
greater efficiency when this handedness is the same as that of the
angular momentum carried by the driving beam.

Fourthly, we can note another factor that is to some extent a
generalisation of the second result from symmetry above. Since, for any given
particle, there are always more modes available with low $|m|$ than with
high $|m|$, scattering from a given $m_0$ to modes with $m = \pm \delta$
will usually be such that modes with $|m| < |m_0|$ will receive more power
than those with $|m| > |m_0|$. As a result, an arbitrary particle placed
in illumination carrying angular momentum, but otherwise arbitrary, will
usually experience a torque.

Therefore, we can summarise some general principles for the
design of optically-driven micromachines:
\begin{itemize}

\item Size matters! The maximum angular momentum available from the beam
is proportional to the radius of the particle, assuming that the
entire beam can be focussed onto the particle \citep{courtial2000}. If
the beam is larger, the portion that misses the particle cannot contribute
to the torque. However, as well as maximising the overlap between the
beam and particle, it is also important to maximise the angular momentum
content of the beam. This can be achieved in either of two ways; firstly,
by choosing a wavelength such that the diffraction-limited Gaussian spot
of a circularly polarised beam is the same size as the particle (since the
angular momentum flux is $P/\omega$, reducing the wavelength maximises
the angular momentum), and secondly by exploiting orbital angular momentum.
The use of orbital angular momentum maximises the angular momentum content
\emph{at a particular wavelength}, allowing a single set of optical
components to be used for micromachines of various sizes. We can also
note that the viscous drag on a rotating sphere is proportional to $r^3$.
Thus, the highest rotation speeds will typically result from smaller
micromachines. The torque, on the other hand, will increase with
increasing size.

\item The rotational symmetry of the particle can be chosen so as to
optimise the torque. The ideal choice depends on the angular momentum
of the incident beam---for an incident beam of azimuthal order $m_0$,
$p$th-order rotational symmetry with $p = m_0$ appears to be a good
choice, allowing coupling of the incident beam to the lowest angular
momentum modes possible, with $m = 0$. More generally,
$m_0 \le p \le 2m_0$ should give good performance. For smaller $p$,
the difference between the magnitudes of the $n = \pm 1$ orders is
less than $2 m_0$, and thus the difference in coupling is likely to
be smaller, and for greater $p$, all scattered orders have a magnitude
of their angular momentum greater than that of the incident beam, which
is likely to increase coupling to the non-torque-producing zeroth
order scattered modes ($n = 0$).

\item Chiral particles can be produced, allowing rotation even by
incident beams carrying little, or even no, angular momentum. Greater
torque will also result, compared with a similar particle,
when illuminated by high angular momentum light of the appropriate
handedness. The price that is paid is that rotation in one direction
is preferred---if one desires equal performance in both directions,
then an achiral device is necessary.

\item The coupling between incident and scattered VSWFs is essentially
a vector version of the coupling between an incident
paraxial Laguerre--Gauss mode and diffracted modes due to a
hologram---in both cases, the phase variation is of the
form $\exp(\mathrm{i}m\phi)$,
and the effect of the symmetry of the particle or hologram on the coupling
is the same. This allows a simple conceptual model of optically-driven
micromachines: microholograms.

\end{itemize}

\begin{figure}[!tbh]
\centerline{%
\includegraphics[width=0.45\textwidth]{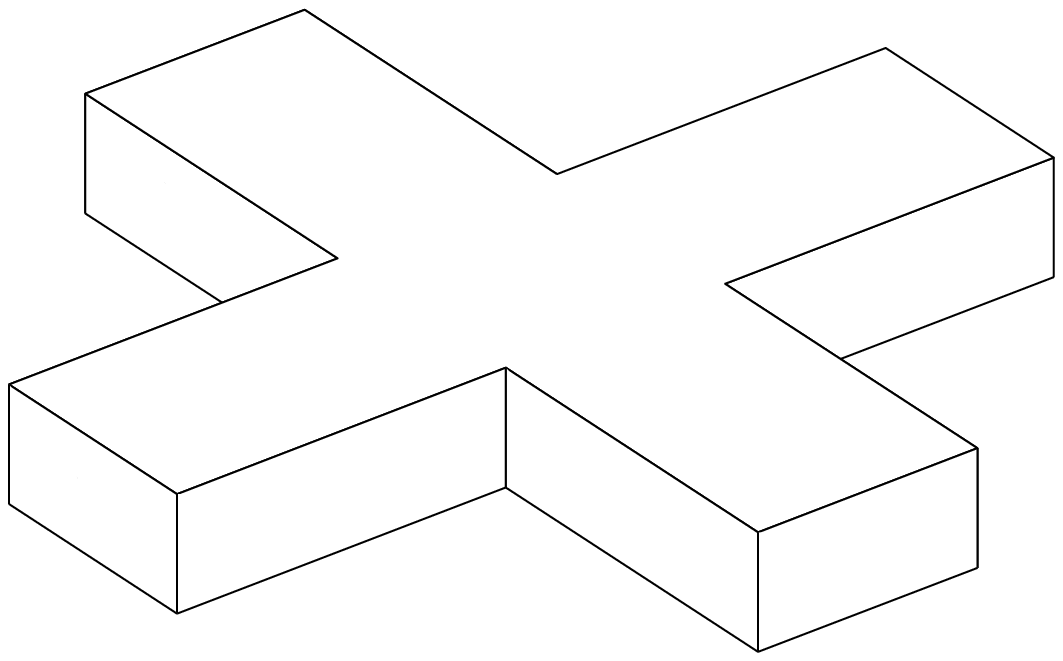}
\hfill
\includegraphics[width=0.45\textwidth]{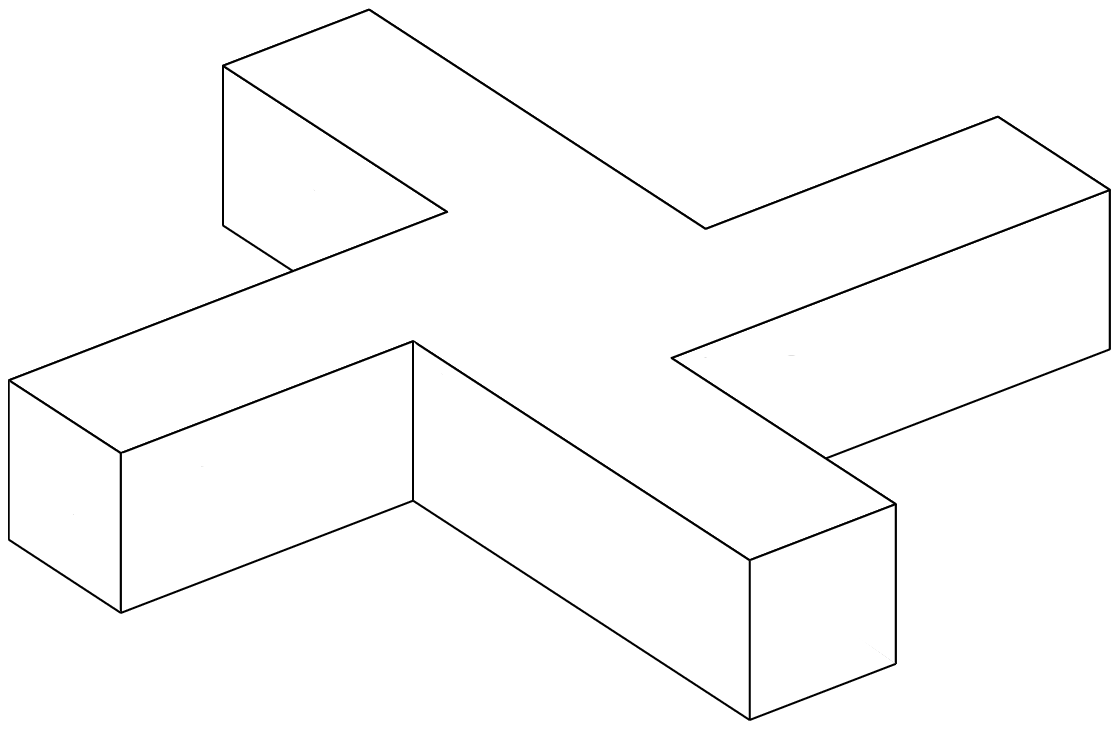}}
\caption{Simple designs for achiral (left) and chiral (right)
optically-driven micromachines, based on the design principles
outlined here. They can be considered as binary phase holograms.
The centres are sold for structural integrity.}
\label{achiral_chiral}
\end{figure}

The principles outlined above have been applied to the design of
optically-driven micromachine elements, and have proved successful
\citep{knoner2007}.
The principles can also be recognised in successful devices
that appear to be based on geometric optics principles, or even
trial-and-error \citep{nieminen2006a}. 
Figure \ref{achiral_chiral} shows two very simple
$p=4$ devices based on these principles. The achiral cross rotor
is designed to be rotated by an incident beam carrying angular
momentum, while the chiral rotor will rotate in an incident
linearly polarised Gaussian beam. Viewing these structures
as microholographic elements, they can be seen as binary phase
approximations of interference patterns between paraxial
Gaussian and LG$_{04}$ beams. If both have planar wavefronts,
the achiral rotor results, while if the Gaussian beam has
curved wavefronts, the chiral rotor results. This suggests
that the direction of rotation of the chiral rotor can be
reversed by reversing the curvature of an incident Gaussian
beam; this effect has been oberved by \citet{galajda2002b}.
Simple structures such as those shown in figure \ref{achiral_chiral}
have also been fabricated and tested \citep{ukita2002}.
The microhologram picture of these devices suggests that the
optimum thickness is that which produces a half-wave phase
difference between light that passes through the arms of
the structure, and light that passes between them.

\section{Optical measurement of optical torque}

One disadvantage resulting from the use of orbital angular
momentum to drive rotation is that while spin angular
momentum is relatively easy to measure optically
\citep{crichton2000,nieminen2001jmo,bishop2003,bishop2004,laporta2004},
it is rather more difficult to optically measure orbital
angular momentum. In principle, it should be possible to
do so by a variety of methods, such as using holograms as
mode filters \citep{mair2001,parkin2004}, interferometry using
Dove prisms to introduce a phase shift \citep{leach2002}, or measurement
of the rotational frequency shift \citep{atkinson1935,basistiy2003}.
While such methods have been demonstrated to be able to accurately
measure optical torque \citep{parkin2004}, accurate optical measurement
of orbital torque within optical tweezers has proved elusive.
This appears to largely result from sensitivity to alignment, both
transverse and angular, and aberrations.

However, a method of estimating the orbital torque has
been demonstrated \citep{parkin2006b,parkin2006c}, wherein
the rotation rate and spin torque are measured for the
same object being rotated by left- and right-circularly
polarised beams, and a linearly polarised beam,
all carrying the
same orbital angular momentum. If the spin torque is
similar in magnitude (the handedness will be opposite) for
the two circularly polarised cases, and the difference
in rotation rate as compared with the linearly polarised
beam is also the same for the two, then we can safely assume
that, as a reasonable approximation, the orbital torque is
the same in all cases. The variation in rotation rate with
the spin then allows the viscous drag to be measured. The
orbital torque can then be estimated from the rotation
rate in the linearly polarised beam.

It should be recalled that for particles with
higher-order rotational symmetry
($p>2$), the torques resulting from the incident left
and right circularly polarized modes are independent of
each other. Furthermore, such particles do not possess
(linear) shape birefringence, and will therefore not
alter the polarization of an incident linearly polarized
beam. Therefore, the spin torques per photon from each of the
left and right circularly polarized components of the beam
must be equal in magnitude and opposite in direction;
if the incident beam is linearly polarized, these torques
must cancel.

So, although at first consideration, the above method for
the measurement of the total angular momentum might strike the
reader as a hopeful approximation, it appears that it is
very likely to yield the correct result.

\section{Conclusion}

From general principles of symmetry, we have been able to deduce
a number of features of the generation of optical torque and its
dependence on the geometries of the particle and the beam.
For an elongated or birefringent particle, whether rotated by
plane polarised light, circularly polarised light, or by a beam with
an elliptical focal spot, the shared symmetry results in similar behaviour
in all of these cases.

Particles of higher-order rotational symmetry, such a microfabricated
optically-driven rotors offer angle-independent torque, and the possibility
of higher torques due to exploitation of orbital angular momentum. An
important design choice is whether to use a chiral particle, for higher
efficiency, or a nonchiral particle, for equal performance in either
direction of rotation. The chiral particle has the advantage of rotating
in a Gaussian beam, whereas the achiral particle requires a beam carrying
orbital angular momentum for rotations. In either case,
a particle with $n$th order rotational symmetry is best driven by a
focussed Laguerre--Gauss mode of helicity of $l\approx n/2$.

It has been suggested in the past that the essentially unlimited
angular momentum of optical vortices, with thousands of $\hbar$ per photon
being achievable, can be used for highly efficient optical rotation. This
is not likely to be useful for the optical rotation
of small objects, since such beams necessarily have focal spots hundreds
of wavelengths across (if, however, one wishes to use an optical rotor that
is hundreds of wavelengths across ...).
In general, the best efficiencies result from the use of focal spots
almost the size of the particle, with as much power as possible in the higher
angular momentum modes.

Finally, we can note that a randomly-selected irregular particle
is almost certainly chiral, and should experience an optical
torque when illuminated.
Why, then, is optical torque regarded as unusual, if most randomly-chosen
particles should experience torques? Although a torque might
exist, it might also be very small; if smaller than torques associated
with the rotational Brownian motion of the particle, the optical torque
will not be noticeable. Rotation should be more readily observable in
environments of low viscosity, such as particles in gas. Under such
circumstances, rotation is observed \citep{ehrenhaft1945}, although
convection and thermophoresis may well be the dominant effects. Torques
on random particles can also be maximised by high refractive index,
increasing the reflectivity \citep{khan2006}. However, most naturally
occurring light is predominatly unpolarised, and as such can be
represented as sums of VSWFs with $m = \pm 1$. The same is the case
for linearly polarised light. It is especially in this case that
torques can result most readily in alignment rather than continuous
spinning, which can also mask the presence of an optical torque.





\end{document}